\documentclass[twocolumn,10pt]{article}

\usepackage[super, comma]{natbib}
\usepackage{graphicx}
\usepackage[font=footnotesize]{caption}
\usepackage{sectsty}
\usepackage{amsmath}
\usepackage{amssymb}
\usepackage[T1]{fontenc}
\usepackage{epstopdf}

\sectionfont{\normalsize}
\subsectionfont{\normalsize}
\subsubsectionfont{\normalsize}
\paragraphfont{\normalsize}

\title{\bf A stellar flare-coronal mass ejection event revealed by X-ray plasma motions}

\author{C.~Argiroffi$^{1,2\star}$, F.~Reale$^{1,2}$, J.~J.~Drake$^{3}$, A.~Ciaravella$^{2}$, P.~Testa$^{3}$, \\
R.~Bonito$^{2}$, M.~Miceli$^{1,2}$, S.~Orlando$^{2}$, and G.~Peres$^{1,2}$ \\
\scriptsize $^1$ University of Palermo, Department of Physics and 
Chemistry, Piazza del Parlamento 1, 90134, Palermo, Italy.  \\
\scriptsize $^2$ INAF - Osservatorio Astronomico di Palermo, Piazza del Parlamento 1, 90134, Palermo, Italy. \\
\scriptsize $^3$ Smithsonian Astrophysical Observatory, MS-3, 60 Garden Street, Cambridge, MA 02138, USA.  \\ 
\scriptsize $^\star$ costanza.argiroffi@unipa.it}

\begin{document}

\maketitle

{\bf

Coronal mass ejections (CMEs), often associated with flares\cite{YashiroGopalswamy2009,AarnioStassun2011,WebbHoward2012}, are the most powerful magnetic phenomena occurring on the Sun. Stars show magnetic activity levels up to $10^4$ times higher\cite{WrightDrake2011}, and CME effects on stellar physics and circumstellar environments are predicted to be significant\cite{KhodachenkoRibas2007,AarnioMatt2012,DrakeCohen2013,OstenWolk2015,OdertLeitzinger2017}. However, stellar CMEs remain observationally unexplored. Using time-resolved high-resolution X-ray spectroscopy of a stellar flare on the active star HR~9024 observed with Chandra/HETGS, we distinctly detected Doppler shifts in S\,{\sc xvi}, Si\,{\sc xiv}, and Mg\,{\sc xii} lines that indicate upward and downward motions of hot plasmas ($\sim10-25$\,MK) within the flaring loop, with velocity $v\sim100-400\,{\rm km\,s^{-1}}$, in agreement with a model of flaring magnetic tube. Most notably, we also detected a later blueshift in the O\,{\sc viii} line which reveals an upward motion, with $v=90\pm30\,{\rm km\,s^{-1}}$, of cool plasma ($\sim4$\,MK), that we ascribe to a CME coupled to the flare. From this evidence we were able to derive a CME mass of $1.2^{+2.6}_{-0.8}\times10^{21}\,{\rm g}$ and a CME kinetic energy of $5.2^{+27.7}_{-3.6}\times10^{34}\,{\rm erg}$. These values provide clues in the extrapolation of the solar case to higher activity levels, suggesting that CMEs could indeed be a major cause of mass and angular momentum loss.

}

\vspace{5mm}

Intense stellar magnetic fields are responsible for the so-called stellar magnetic activity\cite{NoyesHartmann1984,WrightDrake2011}, and for the associated highly energetic phenomena occurring in the outer stellar atmosphere. CMEs, the most energetic coronal phenomena, are observed only on the Sun, because their detection and identification needs spatial resolution.

CMEs are closely linked to flares\cite{WebbHoward2012}. In the standard scenario\cite{ShibataMagara2011} flares are driven by impulsive magnetic reconnections in the corona. The released energy is transported along the magnetic field lines and heats the underlying chromosphere, that expands upward at hundreds of ${\rm km\,s^{-1}}$, filling the overlying magnetic structure (flare rising phase). Then this plasma gradually cools down radiatively and conductively (flare decay). The flare magnetic drivers often cause also large-scale expulsions of previously confined plasma, CMEs, that carry away large amounts of mass and energy. Solar observations demonstrate that CME occurrence, mass, and kinetic energy increase with increasing flare energy\cite{YashiroGopalswamy2009,AarnioStassun2011}, corroborating the flare-CME link.

\begin{figure}[!h]
\centering
\includegraphics[width=8.5cm]{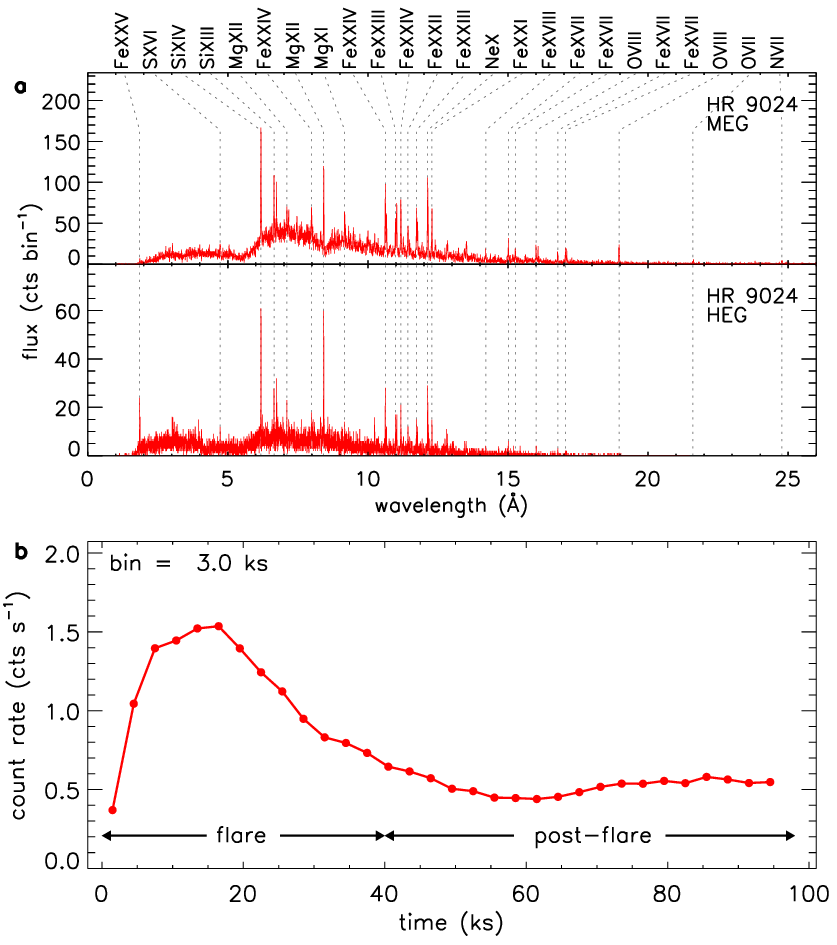}
\caption{Observed X-ray spectra and light curve of HR~9024. {\bf a}, X-ray spectra collected with the Medium Energy Grating (MEG) and High Energy Grating (HEG) during the 98\,ks long {\it Chandra} observation, with the strongest emission lines labeled. MEG and HEG bin size are 5 and 2.5\,m\AA. {\bf b}, X-ray light curve of registered during the {\it Chandra} observation, obtained from the $\pm1$ order spectra of HEG and MEG.}
\label{fig1}
\end{figure}

Active stars have stronger magnetic fields, higher flare energies, hotter and denser coronal plasma\cite{Guedel2004}. Their activity level, measured by the X-ray to bolometric luminosity ratio, $L_{\scalebox{0.6}{X}}/L_{\scalebox{0.6}{bol}}$, can be up to $10^4$ times higher than the solar one. Currently, the properties of stellar CMEs can only be presumed extrapolating the solar flare-CME relation up to several orders of magnitude, even though active stellar coronae differ profoundly from the solar one. This extrapolation suggests that stellar CMEs should cause enormous amounts of mass and kinetic energy loss\cite{AarnioMatt2012,DrakeCohen2013,OstenWolk2015,OdertLeitzinger2017} (up to $\sim10^{-9}\,{\rm M_{\odot}\,yr^{-1}}$ and $\sim0.1L_{\scalebox{0.6}{bol}}$, respectively), and could significantly influence exoplanets\cite{KhodachenkoRibas2007}.

Thus far, there have been a few claims of stellar CMEs. Blueshifted components of chromospheric lines were sometimes attributed to CMEs\cite{HoudebineFoing1990,VidaKriskovics2016,VidaLeitzinger2019}, but they could also be explained by chromospheric brightenings or evaporation events\cite{GunnDoyle1994,BerdyuginaIlyin1999}. CMEs were invoked to explain increased X-ray absorption observed during flares\cite{MoschouDrake2017}. However, the simultaneous variations of flare temperature and emission measure\cite{MoschouDrake2017} ($EM$) insinuate that the increased absorption can be a spurious result coming from the limited diagnostic power of low-resolution X-ray spectroscopy, combined with the oversimplified assumptions of an isothermal flaring plasma and a constant quiescent corona. CMEs were candidates to explain transient UV/X-rays absorptions observed in the eclipsing precataclysmic binary V471~Tau\cite{Wheatley1998}. However, such absorptions can equally be produced by a stellar wind\cite{MullanSion1989}. In addition to their ambiguous interpretation, all these candidate detections never provide the CME physical properties, unless substantial assumptions are made.

We present here strong evidence for the detection and identification of a stellar CME, the estimate of the its properties, and the simultaneous monitoring of the associated flare energetics. The Sun indicates that a flare-CME event should produce hot plasma moving upward and downward within the flaring loop, and, after the flare onset, cool plasma in the CME moving upward. Therefore, monitoring the plasma velocity at different temperatures during a stellar flare is a potentially powerful method of searching for CMEs. The unrivalled X-ray spectral resolution of the Chandra/HETGS jointly with detailed hydrodynamic (HD) modeling allowed us to investigate a very favorable case: the strong flare\cite{TestaReale2007} observed on the active star HR~9024.

HR~9024 is a G1\,III single giant\cite{BorisovaAuriere2016}, with $M_{\star}\sim2.85\,M_{\odot}$ and $R_{\star}\sim9.45\,R_{\odot}$, located at 139.5\,pc. Its convective envelope and rotational period\cite{StrassmeierSerkowitsch1999} (24.2\,d) indicate that an efficient dynamo is at work\cite{BorisovaAuriere2016}, as expected in single G-type giants\cite{PizzolatoMaggio2000}. Even if some contribution to its magnetic field could have fossil origin\cite{BorisovaAuriere2016}, HR~9024 indeed shows coronal\cite{TestaReale2007} ($L_{\scalebox{0.6}{X}}\sim10^{31}\,{\rm erg\,s^{-1}}$, with $T\sim1-100$\,MK) and magnetic field properties\cite{BorisovaAuriere2016} (a dominant poloidal field with $B_{\rm max}\sim10^{2}$\,G) analogous to that of other active stars. Therefore, irrespective of the origin of some magnetic components, and bearing in mind that active stars may show diverse magnetic configurations, the coronal phenomena occurring on HR~9024 can be considered as representative of those of active stars.

HR~9024 X-ray spectrum was collected during a 98\,ks-long Chandra/HETGS observation (Fig.~\ref{fig1}{\bf a}), during which a strong flare (peak luminosity $\sim10^{32}\,{\rm erg\,s^{-1}}$ and X-ray fluence $\sim10^{36}\,{\rm erg}$) was registered (Fig.~\ref{fig1}{\bf b}).

The high energy of this flare maximizes the probability to have an associated CME\cite{YashiroGopalswamy2009}. The flaring loop located near the stellar disk center, as hinted by the Fe fluorescence\cite{TestaDrake2008}, maximizes the possibility to detect radial velocities of both flaring and CME plasmas.

\begin{figure*}[p]
\centering
\includegraphics[width=17cm]{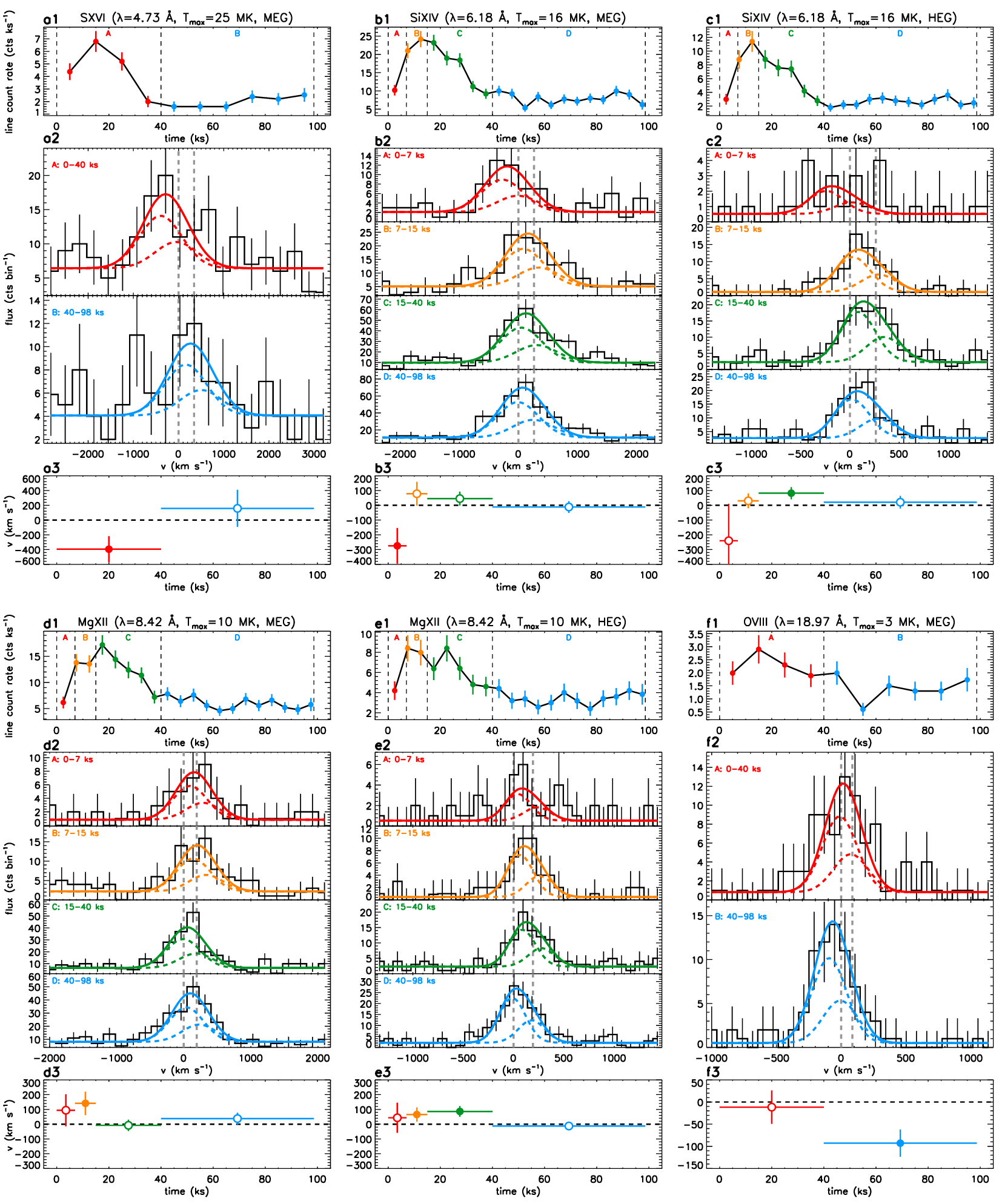}
\caption{Time-resolved line fits. {\bf a}, Analysis of the S\,{\sc xvi} line at 4.73\,\AA, as registered with the MEG $\pm1$ orders. In all plots vertical bars indicate errors at $1\sigma$. {\bf a1}, count rate detected in a 0.1\,\AA\, interval centered on the line. Letters and colors indicate the different time intervals considered to collect the line profile. {\bf a2}, Observed line profile, in different time intervals (black), with superimposed (in different colors, following the same color-code used for the different time intervals) the corresponding best fit function (with dashed curves corresponding to the two transitions of the Ly$\alpha$ doublet, solid curve to their sum). Vertical dashed gray lines mark the rest position of the two Ly$\alpha$ components. On the $x$ axis we report the velocity in the stellar reference frame with respect to the bluest Ly$\alpha$ component. {\bf a3}, Line Doppler shifts, in the different time intervals, computed in the stellar reference frame. Horizontal bars specify the time interval. Filled circles mark velocities different from zero at 1$\sigma$ level at least, open circles indicate velocities compatible with zero. The other plots analogously show the time-resolved analysis of: Si\,{\sc xiv} line at 6.18\,\AA\, as registered with MEG ({\bf b1-b3}) and HEG ({\bf c1-c3}), Mg\,{\sc xii} line at 8.42\,\AA\, as registered with MEG ({\bf d1-d3}) and HEG ({\bf e1-e3}), O\,{\sc viii} line at 18.97\,\AA\, as registered with MEG ({\bf f1-f3}).}
\label{fig2}
\end{figure*}

We measured time-resolved individual line positions, considering only strong and isolated lines that probe plasma with $T$ ranging from 2 to 25\,MK (Fig.~\ref{fig2} and Methods and Supplementary Table~1). We found:

\begin{itemize}
\item[-] significant blueshifts during the rising phase of the flare in the S\,{\sc xvi} line at 4.73\,\AA\, ($-400\pm180\,{\rm km\,s^{-1}}$) and in the Si\,{\sc xiv} line at 6.18\,\AA\, ($-270\pm120\,{\rm km\,s^{-1}}$), with a 99.99\% combined significance of the two line shifts.

\item[-] significant redshifts in the Si\,{\sc xiv} line at 6.18\,\AA\, ($140\pm80\,{\rm km\,s^{-1}}$) and Mg\,{\sc xii} line at 8.42\,\AA\, ($70\pm50$ and $90\pm40\,{\rm km\,s^{-1}}$), during the maximum and decay phases of the flare, with a 99.997\% combined significance of the three line shifts.

\item[-] a significant blueshift in the O\,{\sc viii} line at 18.97\,\AA\, ($-90\pm30\,{\rm km\,s^{-1}}$), after the flare, significant at 99.9\% level.
\end{itemize}

\begin{figure*}[t]
\centering
\includegraphics[width=17cm]{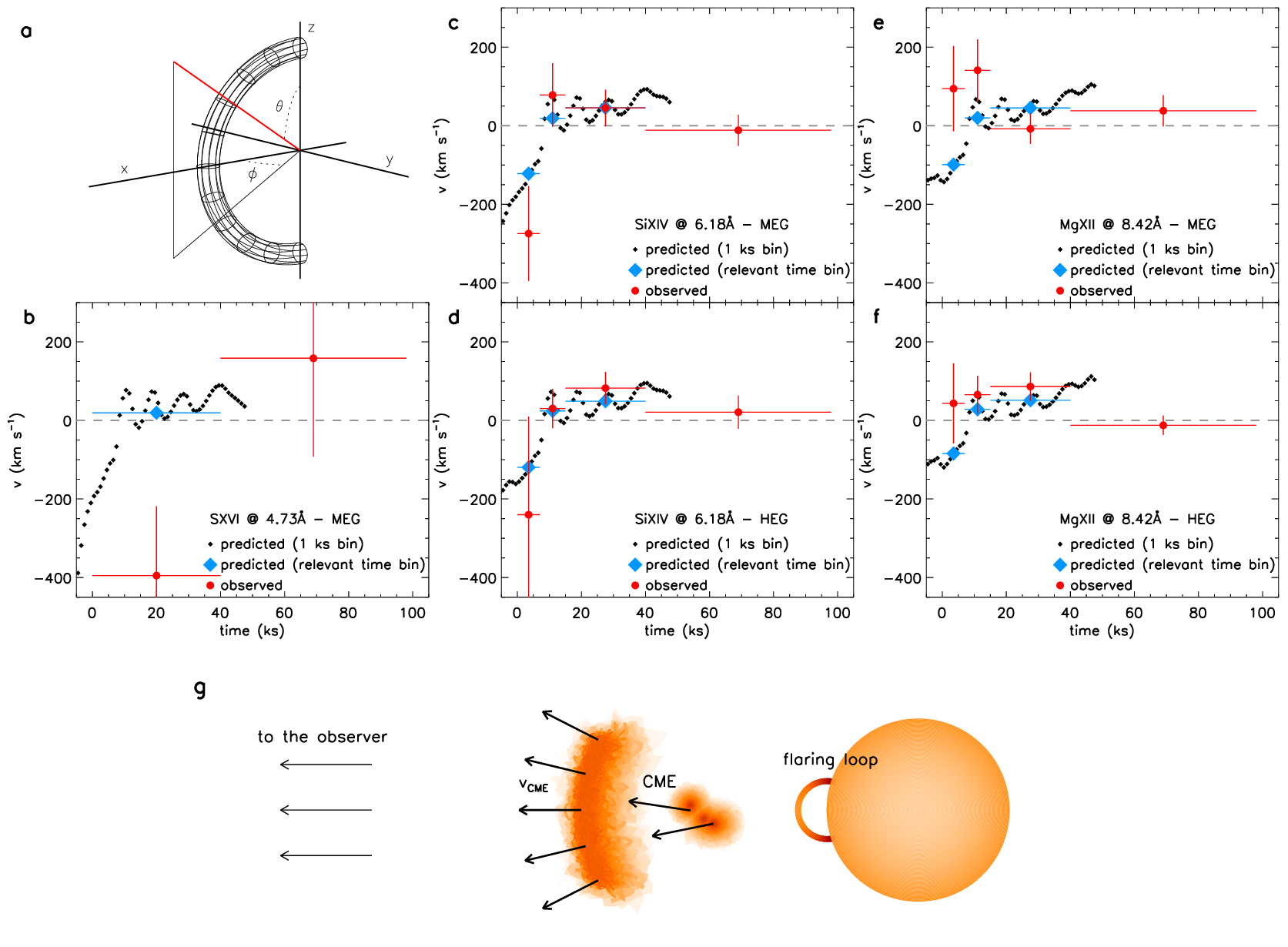}
\caption{Comparison between observed and predicted velocities. {\bf a}, Viewing geometry of the flaring loop with respect to the observer direction (red line) identified by the $\phi$ and $\theta$ angles. {\bf b-f}, Comparison between the line Doppler shifts observed (red circles) and those predicted by the flare model, for $\phi=0^{\circ}$ and $\theta=90^{\circ}$, for the lines showing significant shifts. Small dark diamonds indicate velocities predicted in 1\,ks intervals. Large blue diamonds mark predicted velocities computed integrating the model spectra over the same time intervals considered for the observed spectra. Vertical bars indicate errors at $1\sigma$. Horizontal bars mark to the time interval duration. {\bf g}, Schematic illustration of the star/loop/CME configuration.}
\label{fig3}
\end{figure*}

The first two Doppler shifts tell us that hot flaring plasma moves upward at the beginning of the flare, and then settles back down to the chromosphere, as predicted\cite{ShibataMagara2011}. 
We compared the observed velocities with predictions based on the flaring loop model\cite{TestaReale2007} (Methods and Supplementary Fig.~1). For each line and each time interval, we computed the radial velocity corresponding to the predicted line centroid, for different possible inclinations of the flaring loop (identified by the $\phi$ and $\theta$ angles defined in Fig.~\ref{fig3}{\bf a}). 

The best agreement between observed and predicted velocities (Fig.~\ref{fig3}{\bf b}-{\bf f}) is obtained assuming a loop observed from above ($\phi=0^{\circ}$ and $\theta=90^{\circ}$, see Methods), confirming previous independent results\cite{TestaDrake2008}. The agreement for the Si\,{\sc xiv} and Mg\,{\sc xii} lines is striking (Fig.~\ref{fig3}{\bf c}-{\bf f}). The observed S\,{\sc xvi} blueshift, associated with chromospheric plasma upflows, is of the same order but even more extended in time than expected. The agreement obtained for flaring plasma velocities is an important validation of the standard flare model for flare energies up to $10^{36}$\,erg.

Interestingly, we found that the coolest inspected line, O\,{\sc viii} Ly$\alpha$, that forms at $\sim3$\,MK, is significantly blueshifted (with $v=-90\pm30\,{\rm km\,s^{-1}}$) in the post-flare phase, while no shift is observed during the flare. The low rotational velocity of HR~9024 ($v \sin i \approx 20 \,{\rm km\,s^{-1}}$)\cite{StrassmeierSerkowitsch1999} excludes that this motion is due to structures fixed on the stellar surface.

\begin{figure*}[t]
\centering
\includegraphics[width=17cm]{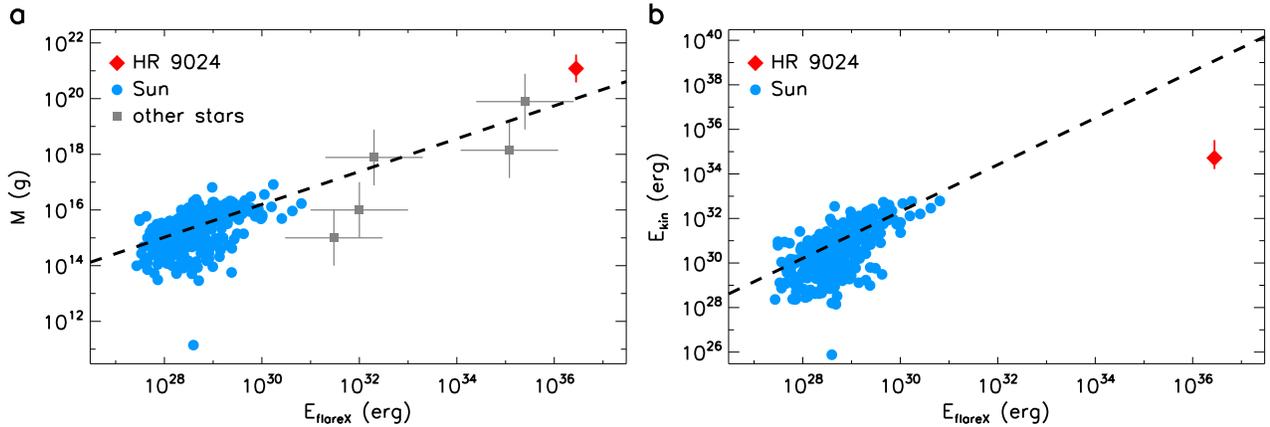}
\caption{Extrapolation of the solar flare-CME relation. Mass $M$ ({\bf a}) and kinetic energy $E_{\rm kin}$ ({\bf b}) of solar CMEs, as a function of the X-ray fluence of the associated flares\cite{YashiroGopalswamy2009}, with the corresponding power-law relations\cite{DrakeCohen2013}, compared to the analogous properties of the observed flare-CME pair on HR~9024, and that of other candidate stellar CMEs\cite{OdertLeitzinger2017}.}
\label{fig4}
\end{figure*}

We identify this motion as the signature of a CME (Fig.~\ref{fig3}{\bf g}): it involves only cool plasma, it occurs after a strong flare located near the stellar disk center, and the observed velocity is within the range of solar CME velocities\cite{WebbHoward2012} (i.e. $20-3000\,{\rm km\,s^{-1}}$). Solar flares are sometimes followed also by the formation of expanding giant arches\cite{Svestka1996}. However, this expansion is very slow ($\sim1-10\,{\rm km\,s^{-1}}$) and possibly only apparent, because due to the sequential brightening of different loops\cite{WestSeaton2015}, and hence unrelated to Doppler shifts. Although the extrapolation of the magnetic configuration of the Sun to more active stars is not straightforward, a CME remains the most rational and only explanation.

Hypothesizing that the CME moves exactly along the line of sight, the distance traveled by the CME in the post-flare phase is $\sim0.8\,R_{\star}$. Most likely the CME started its motion simultaneously to the flare onset (solar flare and CME onsets differ at most of 1\,ks, and the O\,{\sc viii} profile during the flare appears broad, with some blueshifted excess, Fig.~\ref{fig2}{\bf f}). Assuming a constant velocity, the total distance traveled by the CME is $\sim1.4\,R_{\star}$. At this distance from the stellar surface the escape velocity is $220\,{\rm km\,s^{-1}}$, larger by a factor of $\sim2.4$ than the CME velocity. The real CME velocity, as well as its traveled distance, can be higher because of the possible inclination between the CME trajectory and the line of sight. Assuming an inclination of $45^{\circ}$ (i.e. the maximum separation angle for solar flare-CME pairs\cite{YashiroGopalswamy2009,AarnioStassun2011}) the ratio between the local escape velocity and the real CME velocity would reduce to $\sim1.6$.

The initial CME mechanical energy has however a minor importance in determining the CME ending. Solar CMEs follow non-ballistic motions. Magnetic forces and wind interactions often cause strong outward acceleration up to heights of several solar radii\cite{WebbHoward2012}. The detected CME on HR~9024 shows indeed a velocity approximately constant during the post-flare phase ($-100\pm50$ and $-80\pm50\,{\rm km\,s^{-1}}$ in the $40-70$ and $70-98$\,ks time intervals, respectively), indicating that strong magnetic forces act on the CME balancing the stellar gravity. Having no data afterwards, we cannot firmly conclude whether the CME does eventually escape to infinity.

Assuming that the post-flare O\,{\sc viii} line entirely comes from the CME, and inferring a CME temperature of $4\pm1$\,MK, we derive an $EM$ of $(2.8\pm1.0)\times10^{53}\,{\rm cm^{-3}}$ (see Methods and Supplementary Table~2). As in hot plasmas of solar CMEs, we expect that the CME plasma is optically thin and that, in the observed interval, there is no significant heating source\cite{LandiRaymond2010}. The duration of the observed CME X-ray emission (stable in the post-flare phase) indicates a radiative cooling time $\tau>60$\,ks. Being conservative, we assumed $\tau\sim200$\,ks and a factor of 10 for its confidence interval (i.e. $60\,{\rm ks}<\tau<600\,{\rm ks}$). From that we inferred the CME density ($n_{\scalebox{0.6}{e}}$), volume ($V$), mass ($M$), and kinetic energy ($E_{\scalebox{0.6}{kin}}$, see Methods and Supplementary Table~2, obtaining:

\[
n_{\scalebox{0.6}{e}} = (5.5^{+11.8}_{-3.7})\times10^{8}\,{\rm cm^{-3}},
\]
\[
V = (1.1^{+10.3}_{-1.0})\times10^{36}\,{\rm cm^{3}},
\]
\[
M = (1.2^{+2.6}_{-0.8})\times10^{21}\,{\rm g},
\]
\[
E_{\scalebox{0.6}{kin}} = (5.2^{+27.7}_{-3.6})\times10^{34}\,{\rm erg}.
\]

Computing also the X-ray flare fluence ($E_{\scalebox{0.6}{flareX}}\approx 2.8\times10^{36}\,{\rm erg}$), we can compare this flare-CME pair with the solar ones (Fig.~\ref{fig4}). The inferred CME mass suggests that the correlation with the flare energy, observed for the Sun\cite{AarnioMatt2012,DrakeCohen2013}, might hold also for stronger flares at higher activity levels. Despite their unsettled identification (as CMEs or chromospheric evaporation), unconstrained mass estimate (the reported values are lower limits\cite{OdertLeitzinger2017}), and lack of X-ray coverage (X-ray flare fluence was assumed to be comparable to that in the $U$ band\cite{OdertLeitzinger2017}), the  previous candidate stellar CMEs\cite{OdertLeitzinger2017} (gray squares in Fig.~\ref{fig4}{\bf a}) also agree with this high-energy extrapolation. Conversely, the obtained CME kinetic energy is $\sim10^4$ less than expected from solar data extrapolations (Fig.~\ref{fig4}{\bf b}).

Therefore, CMEs of active stars may not be a scaled version of solar CMEs. In terms of the amount of mass ejected, the CME formation mechanism appears to scale smoothly from the solar case to higher flare energy and higher magnetic activity level. In terms of kinematics, remembering that solar CMEs can experience acceleration both in the low corona ($<2\,{R_{\odot}}$) or at large distances\cite{WebbHoward2012}, the CME acceleration mechanism appears instead less efficient, possibly suggesting a different energy partition in the flare-CME pair.

These CME parameters (i.e. a mass compatible with solar extrapolations, and a significantly reduced $E_{\scalebox{0.6}{kin}}$) fit well with magnetohydrodynamic models\cite{DrakeCohen2016,AlvaradoGomezDrake2018}, indicating that the balance between the magnetic forces acting on the CME can be different on active stars, with a higher efficiency of the inward force due to the magnetic tension of the overlying large-scale field with respect to the outward force due to the magnetic pressure of the flux rope.

As integrated effect, assuming that the observed CME eventually escapes the star, the inferred large CME mass supports the hypothesis that CMEs can be a major cause of mass and angular momentum loss in active stars\cite{AarnioMatt2012,DrakeCohen2013,OstenWolk2015,OdertLeitzinger2017}, even if it remains unclear down to what energy flares can cause eruptions in active stars. The diminished $E_{\scalebox{0.6}{kin}}$ to $E_{\scalebox{0.6}{flareX}}$ ratio indicates instead that, at high activity levels, the energy fraction carried out by CMEs diminishes. That supports that magnetic activity at most extracts $\sim10^{-3}$ of the stellar bolometric luminosity, and excludes the huge magnetic energy budget implied by solar case extrapolations to higher activity levels\cite{DrakeCohen2013}.


\section*{Methods}

\subsection*{Data analysis}

HR~9024 was observed on August 2001 for 98\,ks with Chandra/HETGS (ObsID~1892). This instrument configuration consists of two transmission gratings, the HEG and the MEG, used with the ACIS-S detector. The two gratings simultaneously collect spectra in the $1.2-15$\,\AA\ (HEG) and $2.5-31$\,\AA\ (MEG) intervals, with a spectral resolution (FWHM) of 12 and 23\,m\AA, respectively. In this work we inspected the HEG and MEG spectra of HR~9024 separately, each one obtained by adding $+1$ and $-1$ diffraction orders. The data have been retrieved from the Grating-Data Archive and Catalog\cite{HuenemoerderMitschang2011}.

The wavelength calibration of the Chandra/HETGS, allows velocity measurements down to a few tens of ${\rm km\,s^{-1}}$ by comparing line positions within and between observations, as confirmed by several studies \cite{ArgiroffiDrake2017,BrickhouseDupree2001,ChungDrake2004,IshibashiDewey2006}. Hence the Chandra/HETGS is well suited to measure the velocity of flaring and CME plasmas. 

To search for Doppler shifts we selected a sample of strong and isolated emission lines (Supplementary Table~1). Inspecting isolated lines allows us to avoid line position uncertainties due to line blending. Moreover monitoring individual lines, instead of inspecting the whole spectrum, allows us to probe separately plasma components at different temperatures. The selected line sample is composed of: Lyman series lines of N\,{\sc vii}, O\,{\sc viii}, Mg\,{\sc xii}, Si\,{\sc xiv}, and S\,{\sc xvi}; He-like ion lines of Mg\,{\sc xi}, Si\,{\sc xiii}, and O\,{\sc vii}; and the strong Fe\,{\sc xvii} lines in the $15-17$\,\AA\, range. We did not inspect the Ne\,{\sc ix} and Ne\,{\sc x} lines, because of the severe blendings with Fe lines. The maximum formation temperature of the inspected line sample ranges from 2 to 25\,MK.

We determined the position of each selected line by least-squares fitting its observed profile in different time intervals. The selected duration of the inspected time intervals is aimed at: a) separating the phases corresponding to significantly different predicted velocities (HD model, described below and in Supplementary Fig.~1, predicts high upward velocity during the rising phase, and slower downward motions during the maximum and decay phases, Fig.~\ref{fig3}{\bf b-f}); b) collecting enough counts to perform the line fit. For each line we performed the fit in a small wavelength interval, with width $\sim0.1-0.2$\,\AA, around the line rest wavelength. As best-fit function we assumed a Gaussian plus a constant, to take also the continuum emission into account. The $\sigma$ of the Gaussian was fixed to the predicted value\cite{ArgiroffiDrake2017}. Since the Ly$\alpha$ lines are doublets, we fit their observed profiles with two Gaussians, with relative positions fixed to the predicted wavelength difference, and relative intensity fixed to the predicted value (i.e. $2:1$ in the optically thin emission regime).

The observed line shifts, with respect to the predicted wavelengths, provide velocity with respect to the Chandra satellite reference frame ($v_{\scalebox{0.6}{sat}}$). We assumed that these velocities are the same as that with respect to the Earth, because of the low satellite velocity (at most of $\sim1-2\,{\rm km\,s^{-1}}$ with respect to the Earth). We computed the plasma velocity in the stellar reference frame as $v=v_{\scalebox{0.6}{sat}}+v_{\scalebox{0.6}{E}}-v_{\scalebox{0.6}{$\star$}}$, where $v_{\scalebox{0.6}{E}}$ is the line-of-sight Earth velocity at the epoch of the observation (i.e. $v_{\scalebox{0.6}{E}}=19.9\,{\rm km\,s^{-1}}$ in the heliocentric reference frame), and $v_{\scalebox{0.6}{$\star$}}$ is the radial velocity of HR~9024 (i.e. $v_{\scalebox{0.6}{$\star$}}=-1.6\,{\rm km\,s^{-1}}$ in the heliocentric reference frame). Throughout the paper we indicated outward motions with respect to us (redshifts) with positive radial velocities, and inward motions with respect to us (blueshifts) with negative radial velocities.

The lines showed in Fig.~\ref{fig2}, and discussed in the paper are the ones for which a significant shift was detected in at least one time interval. For lines with $S/N$ ratio high enough, we analyzed both HEG and MEG spectra. Because of the different $S/N$ and spectral resolution between corresponding MEG and HEG spectra (with MEG providing higher $S/N$ but lower spectral resolution than HEG), significant line shifts were sometimes found only in one grating. In these cases, the shift measurements obtained with the two gratings (even if only one was significantly different from zero) were anyhow compatible among themselves.

We display in Supplementary Fig.~2 the observed and predicted shifts for those lines with a $S/N$ high enough to allow time-resolved spectral fitting, for which no significant shift was obtained. For the hottest of these lines (i.e. the Si\,{\sc xiii} at 6.65\,\AA, and the Mg\,{\sc xii} at 7.11\,\AA), the expected radial velocities during the flare evolution are high enough to be detectable with the Chandra gratings. However, the low $S/N$ collected for these lines (total line counts are listed in Supplementary Table~1) avoids precise shift measurements and/or prevents from us exploring time bins short enough. For the coolest of these latter lines (i.e. all the Fe\,{\sc xvii} lines), in addition to the low $S/N$, significantly smaller shifts only in very short time intervals are expected. Notice that the large redshifted velocities of $\sim200\,{\rm km\,s^{-1}}$ expected for the very final part of the flare correspond to phases in which the line fluxes become negligible, because of the vanishing $EM$ of the flaring loop.

\subsection*{Flaring loop model and line emission synthesis}

To infer the expected line Doppler shifts due to plasma motions during the flare, and to constrain the loop orientation with respect to the observer, we considered the flare loop model presented by Testa et al.\cite{TestaReale2007}. This model assumes that: the stellar magnetic field is so strong as to confine the plasma inside single closed magnetic tubes (coronal loops); the footpoints of the flaring loop are anchored to the photosphere; the flaring plasma moves and transports energy only along the field lines, and its evolution can be described with a 1D HD model along the tube. The HD equations for a compressible plasma fluid are solved numerically to obtain the evolution of the plasma density, temperature, and velocity along the loop. The flare is triggered with a strong heat pulse injected inside an initially hydrostatic and relatively cool loop atmosphere which includes a tenuous corona linked to a dense chromosphere. Tuning the model parameters to reproduce the observed evolution of $T$ and $EM$ during the flare, it came out that: the total loop length is $5\times10^{11}$\,cm; the duration and rate of the heat pulse are 15\,ks and $1.2\times10^{33}\,{\rm erg\,s^{-1}}$, respectively, for a total injected energy of $\sim1.7\times 10^{37}$\,erg; the pulse heats the plasma to a maximum temperature of $\sim 150$\,MK and makes it expand from the chromosphere at a maximum speed of $\sim1800\,{\rm km\,s^{-1}}$, which drops rapidly below $400\,{\rm km\,s^{-1}}$ after a few ks since the heat pulse has started. The evolution of velocity, temperature, and $EM$ of the flaring loop are shown in Supplementary Fig.~1.

The X-ray emission of the flaring loop is assumed to be optically thin. Line emissivities were retrieved from the APED database\cite{SmithBrickhouse2001}. We computed line emission from the flaring loop considering both short time intervals (i.e. 1\,ks), to monitor line profiles on time scales corresponding to the characteristic time scales of the flaring loop variability, and long time intervals (i.e. corresponding to that adopted for the observed line profiles) to perform a proper comparison between observed and predicted line shifts.

We computed predicted line profiles for different viewing geometries of the loop, exploring the range $0^{\circ}<\phi<90^{\circ}$ and $0^{\circ}<\theta<90^{\circ}$ (Fig.~\ref{fig3}{\bf a}). The $\phi$ angle determines a global scaling factor in the predicted line shifts, with $\phi\sim0^{\circ}$ corresponding to the largest shifts, and $\phi\sim90^{\circ}$ corresponding to no shift. Taking the possible values of $\theta$ into account, and considering that the footpoint portions of the loop are responsible for the highest emission and highest velocity, (Supplementary Fig.~1), configurations with $\theta\sim90^{\circ}$ maximize the line shifts, while configurations with $\theta\sim0^{\circ}$ minimize the line shifts (causing also some line broadening, because of the simultaneous redshifted and blueshifted contributions originating in the motions of plasma located near the loop apex). We found that only for $\phi=0^{\circ}$ and $\theta=90^{\circ}$ the predicted shifts are large enough to reproduce that observed.

In general, since the coronal part of the loop is most of the flaring volume, filled with upflowing plasma at temperatures $\ge50-100$\,MK, we would expect Doppler shifts in highly ionized lines, like Ca\,{\sc xx} formed at $\sim50$\,MK, but they are not detected\cite{TestaReale2007}. Instead, we detected Doppler shifts, associated with motions of flaring plasma, in lines at $T\sim10-25$\,MK, hence emitted mostly at the loop footpoints. This happens because Doppler shift measurements are best obtained in lines detected with high $S/N$ ratio. The spatial distributions of $T$ and $EM$ along the flaring loop determine that hotter lines, mainly emitted by higher loop portions where the $EM$ is lower, have lower $S/N$ with respect to cooler lines, that come primarily from lower loop portions where the $EM$ is higher.

Finally, the agreement obtained between observed and predicted velocities, consdering that this HD flare model was tuned to match only X-ray flux and plasma temperature, support the hypothesis of a flare occurring in a single loop, and allows us to confirming the loop geometry, the temporal and spatial distribution of heating, and the kinetic energy budget involved in plasma motions.

\subsection*{Cool plasma velocity after the flare}

The blueshifted emission, detected at $3\sigma$ level in the post-flare phases in the O\,{\sc viii} line, is per se robust because of the accuracy of the Chandra wavelength calibration: in a sample of active stars\cite{ArgiroffiDrake2017} the shift displayed by this line is always smaller than $13\,{\rm km\,s^{-1}}$. To further corroborate this detection we anyhow inspected the N\,{\sc vii} Ly$\alpha$ at 24.78\,\AA\, that forms mainly at 2\,MK. This line does not have enough counts to fit its profile. Selecting the post-flare interval, the average position of the photons detected in the $\pm1000\,{\rm km\,s^{-1}}$ interval around its rest position is $-110\pm80\,{\rm km\,s^{-1}}$, thus confirming the cool plasma blueshift at $1.4\sigma$ level. These two simultaneous blueshifts further increase the significance of the detected plasma motion. Summarizing, considering all the inspected lines in the post-flare phase, plasma at $T\lesssim4$\,MK moves upward, hence it is located in the CME, possibly representing its hottest component; conversely, plasma with $T\gtrsim5$\,MK appears motionless, and hence situated in stable coronal structures.

\subsection*{CME parameter estimation}

Assuming that the O\,{\sc viii} emission is entirely due to a CME, we have a direct measurement of the CME average radial velocity, $v$, and of the CME total luminosity in the O\,{\sc viii} line, $L_{\scalebox{0.6}{OVIII}}$, corrected for the interstellar absorption\cite{SinghDrake1996} of $4\times10^{20}\,{\rm cm^{-2}}$. To infer the temperature $T$ of the CME we considered the observed post-flare line ratio between the O\,{\sc vii} resonance line at 21.60\,\AA\, and the O\,{\sc viii} Ly$\alpha$. This ratio provides a lower limit\cite{SmithBrickhouse2001} of 3\,MK for $T$. In addition, the Fe\,{\sc xvii} lines, that form mainly at 5\,MK, do not show any blueshift, indicating that they are produced by coronal loops and not by the CME. We therefore deduced that the CME temperature $T$ should be $4\pm1$\,MK.  We do not detect any significant decline in the O\,{\sc viii} line flux. This suggests that the CME likely moves as a coherent structure, without experiencing significant adiabatic expansion, in spite of the large distance travelled in the related time range.

The line luminosity together with the plasma temperature allow us to determine the CME emission measure $EM$ as:

\[
EM = \frac{ L_{\scalebox{0.6}{O\sc{viii}}} }{ A_{\scalebox{0.6}{O}} G(T) }
\]

\noindent
where $G(T)$ is the line emissivity function (APED database\cite{SmithBrickhouse2001}), and $A_{\scalebox{0.6}{O}}$ is the oxygen abundance (we adopted the abundances inferred from the post-flare emission\cite{TestaReale2007}).

The CME emission is observed for $\sim60$\,ks. Therefore, its radiative cooling time $\tau$ must be longer. We assumed $\tau=200$\,ks, with a confidence interval of a factor 10, that corresponds to $60\,{\rm ks}<\tau<600$\,ks. By also assuming that, during the observed emission, there is no heating source in the CME, that its emission is optically thin, and that the hydrogen to electron density ratio $n_{\scalebox{0.6}{H}}/n_{\scalebox{0.6}{e}}$ is 0.83 (value corresponding to high-temperature plasma with cosmic abundances), we can estimate the CME electron density from:

\[
E_{\scalebox{0.6}{int}} = \frac{3}{2}\left( n_{\scalebox{0.6}{e}}+ n_{\scalebox{0.6}{H}} \right) V k_{\scalebox{0.6}{B}} T
\]

\[
\dot{E}_{\scalebox{0.6}{rad}} = n_{\scalebox{0.6}{e}} n_{\scalebox{0.6}{H}} V \Lambda ( T )
\]

\[
\tau = \frac{E_{\scalebox{0.6}{int}}}{\dot{E}_{\scalebox{0.6}{rad}}} \;\; \Rightarrow \;\;
n_{\scalebox{0.6}{e}} = \frac{3}{2}\left( \frac{ n_{\scalebox{0.6}{e}} }{ n_{\scalebox{0.6}{H}} } +1  \right)
                        \frac{ k_{\scalebox{0.6}{B}} T }{ \tau \Lambda ( T ) }
\]

\noindent
where $E_{\scalebox{0.6}{int}}$ is the CME internal energy, $\dot{E}_{\scalebox{0.6}{rad}}$ is the radiative loss rate, $V$ is the CME volume, and $\Lambda ( T )$ is the radiative loss function per $EM$ unit in the $1-2000$\,\AA\, wavelength interval, computed assuming the plasma emissivities of the APED database\cite{SmithBrickhouse2001}. The estimates for $EM$ and $n_{\scalebox{0.6}{e}}$ finally allow us to derive the volume $V$, mass $M$, and kinetic energy $E_{\scalebox{0.6}{kin}}$ of the CME:

\[
V = \frac{EM}{ n_{\scalebox{0.6}{e}} n_{\scalebox{0.6}{H}} }
\]
\[
M = V n_{\scalebox{0.6}{H}} \, m = \frac{EM}{n_{\scalebox{0.6}{e}}} \, m
\]
\[
E_{\scalebox{0.6}{kin}} = \frac{1}{2}Mv^2
\]

\noindent
where $m$ is the mean mass per hydrogen atom. All the values of the relevant CME parameters are reported in Supplementary Table~2.

In the estimation of the uncertainty of $n_{\scalebox{0.6}{e}}$, $V$, $M$, and $E_{\scalebox{0.6}{kin}}$, we considered only the uncertainty on $\tau$, because it is significantly larger than the uncertainties on $T$ and $EM$. Only in the computation of the upper limit of the confidence interval of $E_{\scalebox{0.6}{kin}}$ we included also a factor related to the possible flare-CME separation angle, in agreement with solar observations\cite{YashiroGopalswamy2009,AarnioStassun2011} that indicate at most separations of $45^{\circ}$. The observed O\,{\sc viii} line does not show significant broadening in the post-flare interval, its width is in fact compatible with the instrumental width. Therefore, velocity dispersion along the line of sight in the CME plasma is expected to be small ($\le100\,{\rm km\,s^{-1}}$), corroborating further the inferred $E_{\scalebox{0.6}{kin}}$ value.

We finally notice that both $M$ and $E_{\scalebox{0.6}{kin}}$ are directly proportional to $\tau$. Therefore, the strict lower limit of 60\,ks, provided by the stable post-flare emission in the O\,{\sc viii}, corresponds to the lower limits of the confidence intervals of $M$ and $E_{\scalebox{0.6}{kin}}$. Conversely, the already large upper limit on $M$ is an a posteriori confirmation on the adopted upper limit on $\tau$. Moreover, the reasonable assumptions made for the $\tau$ confidence interval is also supported by the $n_{\scalebox{0.6}{e}}$ value inferred, neatly compatible with the density observed in solar CMEs\cite{LandiRaymond2010,ChengZhang2012,LandiMiralles2013}.


\section*{Data Availability}

The Chandra dataset analyzed in this work (ObsID~1892) can be accessed from http://cxc.harvard.edu/. The data that support plots and findings of this study are available from the corresponding author upon reasonable request.



\section*{Acknowledgements}

The authors acknowledges modest financial contribution from the agreement ASI-INAF n.2017-14.H.O. 


\section*{Author contributions}
C.A., F.R., J.J.D., A.C., P.T., R.B., M.M., S.O., G.P. contributed to scientific discussion and text writing. C.A. and J.J.D. contributed to analysis of observational data. F.R. contributed to the HD model development, and C.A. to the synthesis of the model line profile.


\section*{Competing interests}
The authors declare no competing financial interests.


\section*{Additional information}

Correspondence and requests for materials should be addressed to C.~Argiroffi.


\vspace{18cm}

\onecolumn

\section*{Supplementary Information}

\setcounter{figure}{0}
\renewcommand{\figurename}{Supplementary Figure}
\renewcommand{\tablename}{Supplementary Table}

\begin{table*}[h]
\caption{Selected emission lines}
\normalsize
\begin{center}
\begin{tabular}{lrlcr@{$\pm$}lr@{$\pm$}l}
\hline\hline
 Index$^a$ & \multicolumn{1}{c}{$\lambda^b$}   & Elem. & $T_{\rm max}^c$ & \multicolumn{2}{c}{$F_{\rm MEG}^d$} & \multicolumn{2}{c}{$F_{\rm HEG}^e$} \\
           & \multicolumn{1}{c}{(\AA)}         &       & (MK)            & \multicolumn{2}{c}{(cts)}           & \multicolumn{2}{c}{(cts)}           \\
\hline
 1a &   4.7274 &  S{\sc XVI} & 25.1 &   66 &    15$^f$  &    21 &    10 \\
 1b &   4.7328 &  S{\sc XVI} & 25.1 & \multicolumn{2}{c}{$\cdot\cdot\cdot$}  &  \multicolumn{2}{c}{$\cdot\cdot\cdot$}  \\
 2a &   6.1804 &  Si{\sc XIV} & 15.8 &  505 &    31$^f$  &   231 &    21$^f$  \\
 2b &   6.1858 &  Si{\sc XIV} & 15.8 & \multicolumn{2}{c}{$\cdot\cdot\cdot$}  &  \multicolumn{2}{c}{$\cdot\cdot\cdot$}  \\
 3  &   6.6479 &  Si{\sc XIII} & 10.0 &  218 &    24 &    83 &    14 \\
 4  &   6.7403 &  Si{\sc XIII} & 10.0 &  185 &    23 &    83 &    14 \\
 5a &   7.1058 &  Mg{\sc XII} & 10.0 &  115 &    23 &    34 &    11 \\
 5b &   7.1069 &  Mg{\sc XII} & 10.0 & \multicolumn{2}{c}{$\cdot\cdot\cdot$}  &  \multicolumn{2}{c}{$\cdot\cdot\cdot$}  \\
 6a &   8.4192 &  Mg{\sc XII} & 10.0 &  345 &    26$^f$  &   228 &    20$^f$  \\
 6b &   8.4246 &  Mg{\sc XII} & 10.0 & \multicolumn{2}{c}{$\cdot\cdot\cdot$}  &  \multicolumn{2}{c}{$\cdot\cdot\cdot$}  \\
 7  &   9.1687 &  Mg{\sc XI} &  6.3 &  136 &    21 &    35 &    11 \\
 8  &  15.0140 &  Fe{\sc XVII} &  5.0 &   62 &    14 &    12 &     7 \\
 9a &  16.0055 &  O{\sc VIII} &  3.2 &   52 &    13 &     4 &     6 \\
 9b &  16.0067 &  O{\sc VIII} &  3.2 & \multicolumn{2}{c}{$\cdot\cdot\cdot$}  &  \multicolumn{2}{c}{$\cdot\cdot\cdot$}  \\
10  &  16.7800 &  Fe{\sc XVII} &  5.0 &   53 &    12 &     7 &     6 \\
11  &  17.0510 &  Fe{\sc XVII} &  5.0 &   55 &    12 &     5 &     6 \\
12  &  17.0960 &  Fe{\sc XVII} &  5.0 &   56 &    12 &     4 &     6 \\
13a &  18.9671 &  O{\sc VIII} &  3.2 &  112 &    14$^f$  &     4 &     6 \\
13b &  18.9725 &  O{\sc VIII} &  3.2 & \multicolumn{2}{c}{$\cdot\cdot\cdot$}  &  \multicolumn{2}{c}{$\cdot\cdot\cdot$}  \\
14  &  21.6015 &  O{\sc VII} &  2.0 &    8 &     8 &  \multicolumn{2}{c}{$\cdot\cdot\cdot$}  \\
15a &  24.7792 &  N{\sc VII} &  2.0 &   21 &     9 &  \multicolumn{2}{c}{$\cdot\cdot\cdot$}  \\
15b &  24.7846 &  N{\sc VII} &  2.0 & \multicolumn{2}{c}{$\cdot\cdot\cdot$}  &  \multicolumn{2}{c}{$\cdot\cdot\cdot$}  \\
\hline
\end{tabular}
\end{center}
\footnotesize
$^a$ Lines indicated with the same index number, but different letters, are the two components of H-like resonance doublets. $^b$ Predicted wavelength from the APED database. $^c$ Temperature of maximum formation of the line. $^d$ Line counts observed in the MEG spectrum during the whole observation. $^e$ Line counts observed in the HEG spectrum during the whole observation. $^f$ Lines for which a significant shift was detected in at least one time interval.
\normalsize
\end{table*}

\begin{figure*}[h]
\vspace{2cm}
\centering
\includegraphics[width=17cm]{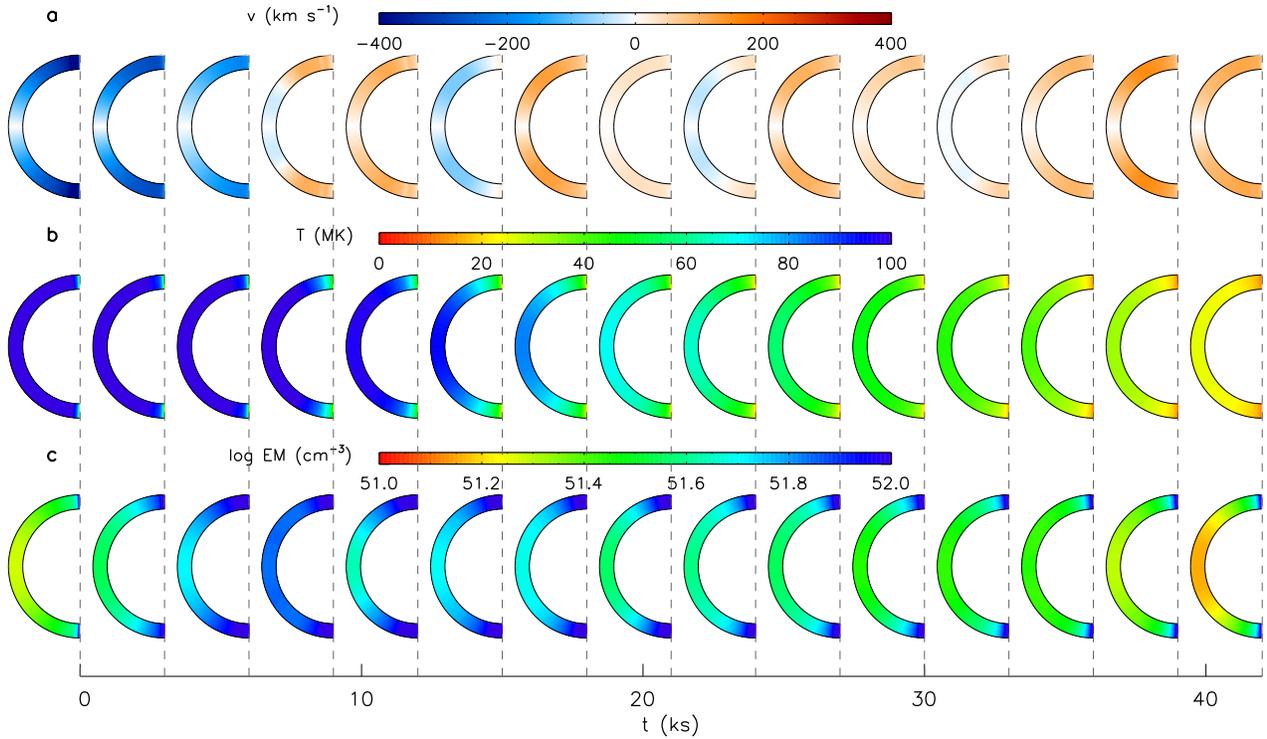}
\caption{Flaring loop model. Time evolution of velocity along the loop ({\bf a}), temperature ({\bf b}), and emission measure ({\bf c}) in the flaring loop, as predicted by the HD model. Negative velocities (marked in blue) correspond to upward motion in the loop, positive velocities (marked in red) correspond to downward motions.}
\vspace{2cm}
\end{figure*}

\begin{table*}
\caption{CME properties}
\normalsize
\begin{center}
\begin{tabular}{llrr}
\hline\hline
parameter & units & value \\
\hline
 $v$ & $({\rm km\,s^{-1}})$ & $ -90 \pm   30$ \\
 $T$ & $({\rm MK})$ & $   4.0 \pm    1.0$ \\
 $M$ & $(g)$ & $(   1.2^{+   2.6}_{-   0.8})\times 10^{      21}$ \\
 $E_{\scalebox{0.6}{kin}}$ & $({\rm erg})$ & $(   5.2^{+  27.7}_{-   3.6})\times 10^{      34}$ \\
 $n_{\scalebox{0.6}{e}}$ & $({\rm cm^{-3}})$ & $(   5.5^{+  11.8}_{-   3.7})\times 10^{       8}$ \\
 $\tau$ & $({\rm ks})$ & $ 200.0^{+ 400}_{- 140}$ \\
 $EM$ & $({\rm cm^{-3}})$ & $(   2.8 \pm    1.0)\times 10^{      53}$ \\
 $V$ & $({\rm cm^{3}})$ & $(   1.1^{+  10.3}_{-   1.0})\times 10^{      36}$ \\
\hline
\end{tabular}
\end{center}
\normalsize
\end{table*}

\begin{figure*}[h]
\vspace{2cm}
\centering
\includegraphics[width=17cm]{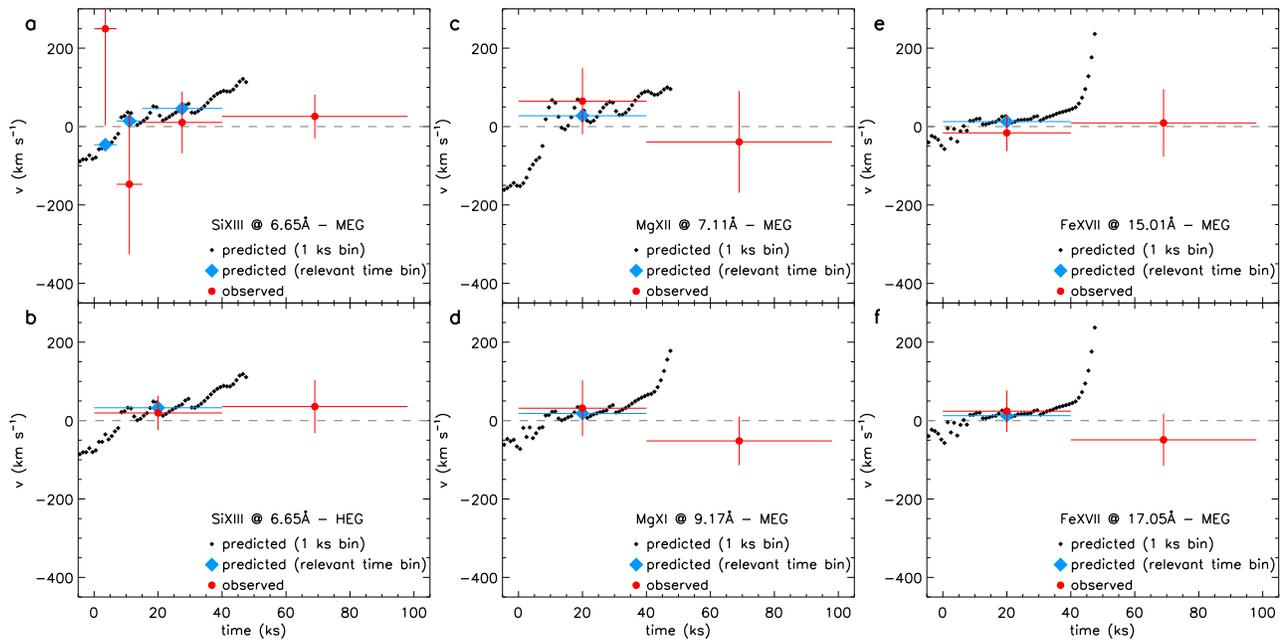}
\caption{Comparison between observed and predicted Doppler shifts for lines showing no significant shifts. Symbols are the same as in Fig.~3.}
\vspace{2cm}
\end{figure*}


\end{document}